\documentclass[12pt,a4paper]{article}

\usepackage{amsmath,amssymb}
\usepackage{epsfig}
\usepackage[dvips]{color}

\newcommand{\bu}{$\bullet$\ }

\newcommand\green{\color{green}}
\newcommand{\be}{\begin{equation}}
\newcommand{\ee}{\end{equation}}
\newcommand{\bes}{\begin{subequations}}
\newcommand{\ees}{\end{subequations}}
\newcommand{\bea}{\begin{eqnarray}}
\newcommand{\eea}{\end{eqnarray}}
\newcommand{\bear}{\begin{equation}\begin{array}}
\newcommand{\eear}[1]{\end{array}\label{#1}\end{equation}}
\usepackage{wrapfig}

\def\lb{\linebreak[4]}
\def\ba{$$\begin{array}}
 \def\ea{\end{array}$$}

\newcommand{\fr}[2]{\dfrac{{ #1}}{{ #2}}}

\newcommand{\fn}[1]{\footnote{{#1}}}
\renewcommand{\le}{\leqslant}
\renewcommand{\ge}{\geqslant}

\def\vep{{\varepsilon}}
\newcommand{\epe}{\mbox{$e^+e^-\,$}}
\newcommand{\ggam}{\mbox{$\gamma\gamma\,$}}

\def\cl{\centerline}

{\end{list}}
\newcounter{enumct}

\newsavebox{\fmbox}
\newenvironment{fmpage}[1]
{\begin{lrbox}{\fmbox}\begin{minipage}{#1}}
{\end{minipage}\end{lrbox}\fbox{\usebox{\fmbox}}}

\begin{document}
%\date{}
%\today
\date{}
\title{Multi-Higgs models. Perspectives for identification of wide set of models in future experiments  at colliders in the SM-like scenario}
\author{I.~F.~Ginzburg
\\
{\small Sobolev Institute of Mathematics, Novosibirsk, 630090, Russia;}\\{\small Novosibirsk State University, Novosibirsk, 630090, Russia}\\
e-mail: Ginzburg@math.nsc.ru\\
{\it Talks given at}\\ {\it Int. Workshop on Multi-Higgs Models.}\\
{\it Lisbon, Portugal 2-5/09/2014}\\{\it and}\\ {\it
 Int. Conf.-Session of Sect. of Nucl. Phys. of Phys. Sc. Div. of RAS}\\ {\it  Physics of Fundamental Interactions, Moscow, Russia 17-21/11. 2014
}}

\maketitle

\begin{abstract}
Higgs mechanism of EWSB can be realized with both the well known minimal model and  more complex non-minimal Higgs models.

These non-minimal models contain new  Higgs bosons -- neutral $h_a$ and charged $H_b^\pm$.  Necessary step in the discovery of such model is observation of these additional Higgses. We discuss the potential of such researches at modern and future colliders in the light of recent LHC results,  for wide set of models, including 2HDM as a simplest example.

Our conclusion is rather pessimistic. The discovery of new neutral Higgs bosons at LHC  is in general  a very difficult task.
(Nevertheless,  some favorable values of parameters of the theory can exist, allowing such observation.)  We propose the regular way for the discovery of  models, which consists in  the study of processes with production of charged Higgs bosons (better, at Linear Collider).

\end{abstract}

\section{Introduction}
%%%%%%%%%%%%%%%%%%%%%%%%%%%%%

This report is based on papers prepared together with M. Krawczyk \cite{GK05}--\cite{GinKr15}  and K.~Kanishev \cite{GKan}, preliminary results were published in \cite{GinZhetF}.

We  base on the following interpretation of modern data \cite{125Higgs}.\\
1. Higgs boson $h$ with mass 125 GeV is discovered at LHC.\\
2. Its properties are close to those in minimal Standard Model (SM) \cite{125_2HDM}.

This situation  does not exclude an extended Higgs sector.  Higgs mechanism of EWSB can be realized with both well known minimal model
and  more complex non-minimal Higgs sector, containing new  neutral  Higgs bosons $h_a$ and charged Higgs bosons $H_b^\pm$.

%%%%%%%%%%%%%%%%%%%%%%%%
\bu \ {\bf
The non-minimal Higgs models}  are devised to solve  various physical problems.

In this report we  consider from common point of view a large group of models with more or less standard description of Higgs phenomenon but with richer set of fundamental scalar fields. Those are models with  $n$ fundamental  weak isodoublets, $p_2$ complex weak  isosinglets $S_c$ and $p_1$  real  weak  isosinglets $S_r$:  $nHDM+p_2(HS_cM)+p_1(HS_rM)$. The models of this group are under wide discussion now.

Examples:

$1HDM$ --  model with single Higgs doublet don't allow CP violation and FCNC. We call this model  SM (Standard Model).

$2HDM$  with huge literature -- see e.g. \cite{TDLee}, \cite{GH05}.  At some values of parameters, 2HDM can explain  CP violation, FCNC, etc. At other values of parameters, 2HDM gives Dark Matter (Inert doublet Model) (without CP violation in Higgs sector and FCNC) (see e.g. \cite{inert}). One more set of parameters realizes Higgs sector of MSSM.

$2HDM +1(HS_cM)$ describes Higgs sector of nMSSM (see also \cite{Chen}).

$3HDM$ at suitable set of parameters describes models with Dark Matter and possible CP violation and FCNC \cite{Bogdan,Arando}.

$nHDM$ at $n\ge 3$, in particular $n=6$ \cite{Wacker}, can describe Dark asymmetric Matter (see e.g. \cite{IgorZp}).

These models give $n-1$ pairs of charged Higgs bosons $H_b^\pm$ with masses $M^\pm_b$ and $2n-1+2p_2+p_1$ neutral scalar Higgs particles $h_a$ with masses $M_a$, having either definite or indefinite CP parity (in the latter case we have CP violation).  Variants with suitable Yukawa sector allow to have flavour changing neutral currents (FCNC), etc.

$\lozenge$ We don't discuss models with alternative explanations of situation  or (and) with additional mechanisms, supplementing or changing the standard Higgs mechanism -- little Higgs, orbifold, radion,   models with  Higgs triplets,... We have not  found a general description for this group of models.

\bu \  Necessary step in the discovery of new model is observation of additional Higgses $h_a$, $H^\pm_b$. Such discovery is a  challenge for the next stage of LHC and \epe \ LC.  A number of papers, devoted this task, study different non-minimal models with various benchmark parameters of new particles and (or) parameters of Lagrangian \cite{genref}. They usually find  that many "natural" approaches in these problems turn out either non-realistic or very difficult (for example, demand extremely high luminosity integral).

In this report we show that such conclusion is not  caused by an unfortunate choice variant of non-minimal model or its parameters but is general feature  for all models of the considered class with almost arbitrary parameters, compatible with modern data  (sect.~\ref{secSMsit}). To come to this conclusion we use sets of Sum Rules (sect.~\ref{secSR}) which form either extensions of known in 2HDM Sum Rules for more wide class of models with arbitrary set of parameters or new Sum Rules founded recently. In the SM-like scenario, Sum Rules allow to
describe many properties of new Higgs bosons (sect.~\ref{seccons}). In sect.~\ref{seccouplfin} we discuss  possible values of couplings of new neutrals. In sect.~\ref{secha} we discuss  properties and production of these new neutrals.

In the sect.~\ref{sec2HDM} we consider the important particular case -- the most general 2HDM (with $h_1$ -- discovered Higgs boson and two new neutrals $h_{2,3}$) in SM-like scenario. First, we note that the coupling $Zh_2h_3$ is not small, while couplings $Zh_1h_2$ and $Zh_1h_3$ are small. Second, we consider possible observation  of triple Higgs vertex $h_1h_1h_1$ and show that it is unlikely
to see deviation from SM prediction in this observation
(except exotic cases).

We summarize results in sect.~\ref{secsum}.

%%%%%%%%%%%%%%%%%%%%%%%%%%%%%
\bu \
In the discussion  we   use {\bf  relative couplings}, defined as
\be
\chi^P_{a}=g^P_a/g^P_{\rm SM}\,, \qquad  (P=V\,(W,Z) , q=(t,b,...), \ell=(\tau,...))\,.\label{relcoupldef}
\ee
Here $g^P_a$ is the the couplings of  neutral Higgs boson $h_a$   with the fundamental particle $P$ and $g^P_{\rm SM}$ -- similar coupling in the SM.

The models with charged Higgs bosons contain   vertices $H^+_bH^-_bh_a$ and $H^\pm_b W^\mp h_a$. For them,  we define relative couplings
\begin{equation}
\chi_{a}^{H^\pm_b W^\mp} = \dfrac{g(H^\pm_b W^\mp h_a)}{M_W/v}\,; \quad  \chi^{\pm b}_a=\fr{g(H^+_bH^-_bh_a)}{2M_\pm^2/v}\,.\label{relcoupb}
\ee

The neutrals $h_a$  generally have no definite CP parity. Couplings $\chi^V_a$ and $\chi^{\pm b}_a$ are real due to Hermiticity of Lagrangian, while other couplings  are generally complex.
The $Re(\chi_a^f)$ and $Im(\chi_a^f)$ are responsible for the interaction of fermion $f$ with CP-even and CP-odd components of $h_a$ respectively.

$\lozenge$ In the CP conserving case some of $h_a$ are scalars, others are pseudoscalars. In this case we have
\be
(a)\;\boxed{\prod\limits_a \chi^V_a=0}\,, \quad (b)\; \boxed{\prod\limits_a \chi^{\pm b}_a=0}\,,
\quad (c)\;\boxed{\left|\prod\limits_a \chi^f_a\right| =\prod\limits_a \left|\chi^f_a\right| \;\; \mbox{for each fermion $f$}}\,. \label{CPchi}
\ee
(In the  2HDM with CP conservation we have $h_3=A$ (pseudoscalar) and $\chi^V_3=0$, $\chi_3^\pm=0$, $Im(\chi_{2,1}^f)=0$, $Re(\chi_3^f)=0$. In this model the  relationship (\ref{CPchi}c) follows from the (\ref{CPchi}a).)

We  omit the adjective "relative" further in the text. Besides, we omit superscript $b$ for models with single $H^\pm$.

%%%%%%%%%%%%%%%%%%%%%%%%%%%%%%%%%%
\section{Modern status. SM-like scenario }\label{secSMsit}
%%%%%%%%%%%%%%%%%%%%%%%%%%%%%%%%%%%%%%%%%%%%%%%

The intensive study of recently discovered Higgs boson makes very probable that the {\it SM-like scenario}
 \cite{GKO}, (or {\it  SM alignment limit \cite{Pilaf15}})
 is realized in the nature:\\
1)  Single observed Higgs boson $h$ has mass $M_h\approx 125$~GeV, we denote it as $h_1$.\\
2)  Its couplings to fundamental particles $P$ (gauge bosons $V$ and fermions $f$) are close to the SM  expectations within experimental accuracy (see e.g. \cite{125_2HDM}):
 \be
\vep_P=\left|1-|\chi^P_1|^2\right|\ll 1\,\qquad (P=V(W,\,Z),\;\;\; f=(t,\,b,\,\tau,...))\,.\label{SMlike}
 \ee
However, this statement remains only a plausible hypothesis before we are able to measure the coupling with a sufficient accuracy. In numerical estimates we have in mind strong version of this inequality, with $\vep_V\sim 0.1$ ("strong SM-like situation").

{\it The existence of SM-like scenario don't close the doors for realization of non-minimal Higgs models.} No doubts that the SM-like scenario in the non-minimal model can occur if additional Higgs bosons are very heavy and are coupled only weakly with usual matter (decoupling limit) (see  e.g. \cite{TDLee}).  15 years ago it was found  that, at finite  inaccuracy of future experiments (at both the LHC and the planned high energy $e^+e^-$ collider), even the simplest non-minimal model 2HDM with the special choice of the Yukawa interaction 2HDM-II allows several possible windows significantly differing from the decoupling limit and admitting the SM-like scenario\fn{In refs.~\cite{GKO}
our problem was to understand if  Photon Collider  can distinguish different Higgs models in the case when future LHC and  $e^+e^-$ collider  show no visible deviations from SM.} \cite{GKO}. Naturally, such windows exist in other models as well.

The  future experiments  reduce $\vep_P$ and, consequently, the region of the allowed parameters of each non-minimal model.

%%%%%%%%%%%%%%%%%%%%%%%%%%%%%
\section{Sum Rules}\label{secSR}
%%%%%%%%%%%%%%%%%%%%%%%%%%%%%%

The freedom in  parameters of discussed models is limited by Sum Rules (SR), which are going to be the key point in our discussion. \\

\bu \ {\bf  The   coupling of the  EW gauge boson $V$ to each  neutral Higgs scalar $h_a$ ($\pmb{\chi^V_a}$)} is  real due to Hermiticity of Lagrangian.  In models like nHDM (with $p_i=0$)
$\chi^Z_a=\chi^W_a\equiv \chi^V_a$.

These $\chi^V_{a}$ coincide with elements of rotation matrix, describing transition from neutral components of Higgs fields $\Phi_j$ to the physical neutral Higgs bosons $h_a$ in the Higgs basis.  The unitarity of this transformation matrix results in the  Sum Rules, which are valid for all discussed models both with and without CP violation:
 \be
\sum\limits_a (\chi^W_{a})^2=1\,,\;\; \sum\limits_a (\chi^Z_{a})^2=1:\qquad
 \boxed{\mbox{all Higgs models with doublets and singlets}}\,.\label{SRV}
 \ee
This argumentation spreads  the approach of  \cite{GKan} developed for the most general 2HDM. (For  particular case  of 2HDM  such SR's
were obtained earlier in   \cite{gunion-haber-wudka},\cite{GK05}). One can say that these SR's mean that masses of gauge bosons are given by the single v.e.v. $v$. \\

\bu \ {\bf The  couplings  $\pmb{\chi^f_{a}}$}  of each definite fermion  $f$ (quark or lepton) to all neutral Higgs scalars $h_a$ are generally complex.

These Sum Rules   naturally appear  in  models $nHDM +p_2HSn_2M+p_1HSn_1M$ at arbitrary $n$ and $p_i$, where weak  isosinglets are not coupled with fermions.
To prove this,  we start with Sum Rule proved for 2HDM with  definite Yukawa interaction  (Model I or Model II) in  \cite{gunion-haber-wudka}, \cite{GKO}, \cite{GK05}. Let us write general Yukawa term for interaction of down-type fermion $f$ to neutral components\fn{Similar argumentation is valid for up-type fermion with the change $\phi_{j,0}\to\phi^*_{j,0}$.} $\phi_{j,0}$ of $\phi_j$ as $\Delta L_Y =\sum_j g_{jf}\bar{\psi}^\dagger\phi_{j,0} \psi_f$. Simple reparameterization $\phi_{1,0}^\prime=N\sum_j g_{jf}\phi_{j,0}$ (where $N$ -- the normalization factor) transforms this term to the form $\Delta L_Y = g_{1f}^\prime\bar{\psi}^\dagger\phi_{1,0}^\prime \psi_f$, which coincides with that for Model I (or II) in 2HDM (we call that {\it f-selective reparameterization basis} \cite{GKan}). For the latter case Sum Rules have been proved in \cite{gunion-haber-wudka,Celis} in the form\fn{Another  proof of these SR's is similar to that developed in \cite{GKan} for 2HDM. Couplings $\chi^f_a$ can be expressed via couplings $\chi^V_a$, $\chi^{H^+W^-}_a$ and parameters of transformation of Higgs  basis to the f-selective basis. The orthogonality of this transformation leads to SR's \eqref{SRf}.} \bear{c}
\sum\limits_{a}(\chi^f_{a})^2=1: \;\; \boxed{\mbox{models without  Yukawa interaction with isoscalars $S_i$}}.
\eear{SRf}
Our argumentation shows that these SR's are valid for much more general class of models than those discussed in  \cite{gunion-haber-wudka,Celis}.\\

\bu \ {\bf The  non-diagonal couplings  to EW gauge bosons $\pmb{H^\pm W^\mp h_a}$} \eqref{relcoupb} are generally complex.  The Higgs potential is naturally invariant under {\it rephasing transformation} $\phi_i\to \phi'_ie^{i\alpha_i}$, compensated by the  corresponding phase rotation of some  coefficients.  This rephasing freedom results in the  phase freedom  of couplings $\chi_{a}^{H^\pm W^\mp}\to \chi_{a}^{H^\pm W^\mp} e^{i\beta} $, keeping phase differences between $\chi_{a}^{H^\pm W^\mp}$ for different $a$. The SR's for these quantities were obtained firstly  in  \cite{GKan} for the most  general 2HDM.  The method of derivation of these SR's allows to extend result for all models with single charged Higgs boson  and arbitrary number of Higgs singlets, $2HDM+p_2(HS_cM)+p_1(HS_rM)$,
\bear{c}
|\chi_{a}^V|^2+| \chi_{a}^{H^\pm W^\mp}|^2=1:\;\boxed{\mbox{$2HDM$ +{\it isoscalars}, with arbitrary Yukawa sector}}\,.
  \eear{SRch}

$\lozenge$ {\bf Some relations for different Yukawa sectors in 2HDM}. The Yukawa couplings for different fermions are generally independent on each other. In some widely discussed models of Yukawa sector these couplings are correlated.

For the 2HDM-I, the $f$-selective bases coincide for all fermions, $\beta_t=\beta_b$, $\xi_t=\xi_b$, and
\bes\label{Yuk}\be
\chi^{u}_a=\chi^{d}_a=\chi^{\ell}_a\,.\label{YukI}
\ee

For the 2HDM-II,  the $u$-selective bases coincide for all up-quarks, and the $d$-selective bases coincide for all down quarks, $\beta_b=\pi/2-\beta_t\equiv\beta$, $\xi_b=\xi$, $\xi_t=0$. In this case the pattern relations among Yukawa couplings for different fermions take place \cite{GK05}
\be
(\chi^u_{a} +\chi^d_{a})\chi^V_{a}=1+\chi^u_{a}\chi^d_{a}\,.
\label{YUkII}
\ee\ees

%%%%%%%%%%%%%%%%%%%%%%%%%%%
\section{Consequences from SR's in the SM-like scenario}\label{seccons}
%%%%%%%%%%%%%%%%%%%%%%%%%%

In this section we have in mind usually $a\ge 2$.

%%%%%%%%%%%%%%%%%%%%%%%%%%%
\subsection{Couplings}\label{seccouplfin}
%%%%%%%%%%%%%%%%%%%%

1. Because of \eqref{SRV}, couplings of  neutrals $h_a$ to gauge bosons $\chi^V_a$ are small (these Higgses are {\it gaugefobic}),
  \be
  |\chi^V_a|^2<\vep_V\ll 1\,.\label{chVa}
  \ee

2. Because of \eqref{SRch}, \eqref{SRV}, the absolute values of non-diagonal couplings  to EW gauge bosons $\chi^{W^\pm H^\mp}_a$ are close to their maximal values while similar coupling for the observed Higgs $\chi^{W^\pm H^\mp}_1$ is small:
\bea
a) \;\;|\chi^{W^\pm H^\mp}_a|^2\approx 1\,;\quad b)\;\;
|\chi^{W^\pm H^\mp}_1|^2\sim \vep_V\ll 1\,.\label{chWa}
\eea
(The calculations of $H^-\to W^-h_1$ decay at LHC in \cite{WHh1} are made in the specific case of CP-conserving 2HDM  and, in fact, in the case when strong SM-like scenario is violated.)

3. The SR's for couplings to the given fermion $f$ \eqref{SRf} can be written as $\sum\limits_{a\ge 2} (\chi^f_a)^2\approx 0$.
We will write here about the most important case $f=t$.
Since couplings $\chi^t_a$ are generally complex, this SR can be saturated by different ways, we discuss the simplest limiting cases
\bes\label{fcoupl}
\bea
a)\qquad & |\chi^t_a|<1\quad\mbox{ for all}\quad h_a\,,\label{fsmall}\\
b)\qquad &\!\!\!\!\!\!(bI)\quad |\chi^t_a|\approx |\chi^b_a|\gg 1; \qquad  (bII)\quad |\chi^t_a|\approx |1/ \chi^b_a|\gg 1\,,\label{flargeI_II}\\
c)\qquad &|\chi^t_{a_2}|\approx |\chi^t_{a_1}|>1\,,\quad
\chi^t_{a_2}\approx i\chi^t_{a_1}\quad\mbox{ for some $h_{a_1}$ and $h_{a_2}$}\,. \label{flarge}
\eea\ees

The case \eqref{fsmall} provides no new interesting opportunities in the discovery of $h_a$.

In the opposite case, if some couplings $\chi^t_{a}$ are large in their absolute value, new interesting opportunities appear\fn{The standard perturbative estimates, used here and in other papers, become invalid if this coupling is enormously large, at $|\chi^t_{a_1}|>2\pi$ we come into the region of the strong interaction in Higgs sector, transferred by $t$-quarks.}, depending on variant of  organization of Yukawa sector.

The eq.~\eqref{flargeI_II} describes two limiting options in the organization of Yukawa sector. The case (bI) corresponds to the Yukawa sector similar to that in 2HDM-I. The case (bII) corresponds to the Yukawa sector similar to that in 2HDM-II.

One particular opportunity in saturation of \eqref{SRf} is described by eq.~\eqref{flarge}. In this case  the absolute values of couplings $\chi^t_{a_i}$ are large only for two neutrals $h_a$. For one of these neutrals $Re(\chi^t_{a_i})>Im(\chi^t_{a_i})$, for another neutral the imaginary part dominates. (This variant can be realized in CP conserving 2HDM  with $h_{a_1}=H$, $h_{a_2}=A$.)

The opportunities \eqref{flargeI_II}  and \eqref{flarge} can coexist or not coexist.

%%%%%%%%%%%%%%%%%%%
\subsection{Properties and production of neutral Higgses $\pmb{h_a}$}\label{secha}
%%%%%%%%%%%%%%%%%%%

In all considered models the masses $M_{a>1}$  and $M^\pm_b$ as well as some couplings of neutrals to vector bosons can be treated as independent  free parameters, limited only by SR's  written above.
(The limitations for the masses  can result  from some additional hypotheses, implemented in model.) Some triple and quartic couplings are also independent parameters of theory (see 2HDM \cite{GKan} where this complete set contains masses of Higgs bosons $M_a$, $M_\pm$, two of three couplings $\chi^V_a$, and couplings $g(H^+H^-h_a)$, $g(H^+H^-H^+H^-)$). The Yukawa couplings form additional set of input parameters.

For definiteness, we assume\fn{The opportunity that some neutrals are lighter than 125~GeV cannot be excluded, but this opportunity is strongly constrained  by modern data. } $M_a>150$~GeV and $|\chi^f_a|<40$ for $f\neq t$. To make some statements more transparent, we will compare discussed quantities with those for the would-be SM Higgs boson having the same mass $H^{(wb)}_{SM}(M_a)$, for example total width $\Gamma^{(wb)}_{SM}(M_a)$ and some cross sections, like $\sigma^{(wb)}_{SM}(gg\to h|M_a)$.

Some conclusions below about total width and observability
can be changed by effects of moderately strong interaction in Higgs sector with large triple Higgs vertices like $h_ah_1h_1$, etc. They should be considered separately. Here we neglect this opportunity.

\bu {\bf Effects from coupling $\pmb{h_a}$ to gauge bosons}. For
the $H^{(wb)}_{SM}(M_a)$  with  mass $M_a>150$~GeV
the main contribution to the width  comes from the decays $h\to W^+W^-$ and $h\to ZZ$. These decays and processes like $W$ fusion provide the main signal  for detection of the  Higgs boson.
The production of this  $H^{(wb)}_{SM}(M_a)$  through a
gauge vertex provides the
best signal/background ratio and the least inaccuracy in the measurement of its parameters both at the LHC and at the ILC.

Oppositely,
according to eq.~\eqref{chVa},\\
{\it (i)} $\Gamma_a\ll \Gamma^{(wb)}_{SM}(M_a)$.\\
{\it (ii)} The decay $h_a\to W^+W^-/ZZ$ is suppressed, observation of $h_a$ via this decay is hardly probable.\\
{\it (iii)} The  search for new neutral Higgs bosons in the $W$ fusion at the LHC, $e^+e^-\to Zh_a$ and $e^+e^-\to \nu\bar{\nu}h_a$ at the ILC, and $e\gamma\to\nu W^- h_a$  at the PLC
(photon collider) cannot be successful \cite{GinKr13}, their cross sections are roughly by one order of value lower then those calculated for $H^{(wb)}_{SM}(M_a)$.

\bu The decreasing of partial widths and production cross sections, caused by small coupling to gauge bosons can in principle be compensated by {\bf the interaction $\pmb{h_a\bar{t}t}$ }.

$\lozenge$ {\bf In the case \eqref{fsmall}} such compensation is  absent, that's why  $h_a$ is very narrow resonance with small partial widths and small cross section of gluon fusion. As a result, the observation of this particle occurs to be the very difficult problem.  Besides, the associative production ${gg\to t\bar{t}h_a}$ is suppressed as compare to the SM case.

$\lozenge$ {\bf The cases \eqref{flargeI_II}}.

{\it The two gluon width of Higgs boson} $\Gamma_a^{gg}$ is saturated by contribution of $t$-quark loop. Therefore, this width is enhanced comparing with the would-be SM case by a factor $|\chi^t_a|^2$, just the same as {\it cross section of gluon fusion.}   In the one-loop approximation
\be
\sigma(gg\to h_a)=\sigma^{(wb)}_{SM}(gg\to h|M_a)\left[|Re(\chi^t_a)|^2+|Im(\chi^t_a)|^2\Phi^{(O/E)}(4M_t^2/M_a^2)\right].
\label{gfus}
\ee
Here $\Phi^{(O/E)}(r)$ is the ratio of two well known loop integrals, defined for CP-odd and CP-even Higgs bosons respectively (see e.g. \cite{Ginhhh}), at $M_a=300$~GeV we have $\Phi^{(O/E)}(r)\approx 2.7$.

{\bf At $\pmb{M_a<350}$~GeV} the cross section of $h_a$ production via gluon fusion is given by eq.~\eqref{gfus}, i.e. it is enhanced in comparison with similar cross section for the would-be SM Higgs boson with mass $M_a$.

$\triangledown$ If Yukawa sector is similar to Model I (variant (bI)), we have $\chi^b_a\approx\chi^t_a$. For the $H^{(wb)}_{SM}(M_a)$  we have the $BR^{(wb)}_{SM}(h\to b\bar{b}|M_a>200~GeV)<0.003$, therefore  the $b\bar{b}$ channel for the hunting for such Higgs is practically closed. Oppositely, at  $|\chi^b_a|\approx|\chi^t_a|\gg 1$
one can hope to observe $h_a$ as the  narrow peak in the production of $\bar{b}b$ pairs   at LHC.\\
\begin{fmpage}{0.95\textwidth}
\cl{ {\bf Benchmark example for CP-even $h_2$ with $M_2=300$~GeV}.}
\begin{center}\begin{fmpage}{0.85\textwidth}
\cl{\it Properties of the would-be SM Higgs boson with this mass.}
$\Gamma^{(wb)}_{SM,tot}=8.4$~GeV, $BR^{(wb)}_{SM}(h\to b\bar{b})\approx 0.0008$, $\Gamma^{(wb)}_{SM}(h\to gg)\approx 3.4$~MeV.
\end{fmpage}\end{center}
 Let  $\boxed{|\chi^t_2|=|\chi^b_2|= 6$, $|\chi_2^V|=0.2}$.
In this case\\  $\Gamma(h_2\to b\bar{b})=|\chi_2^b|^2\cdot 7$~MeV $\approx 250$~MeV,\\
$\Gamma(h_2\to W^+W^-(ZZ))=|\chi_2^V|^2\cdot 8.4$~GeV $\approx 340$~MeV,\\
$\Gamma(h_2\to gg)=|\chi_2^t|^2\cdot 3.4$~MeV $\approx 120$~MeV.

It gives $\Gamma_2\approx 0.7$~GeV with $BR(h_2\to b\bar{b})\approx0.36$.

The cross section $\sigma(gg\to h_a\to \bar{b}b)\approx
|\chi^t_a|^2\sigma^{(wb)}_{SM}(gg\to h|M_a)BR(h_2\to b\bar{b})$.\\
The account of CP odd admixture in $h_2$ increases result -- see \eqref{gfus}.

\end{fmpage}\vspace{2mm}

If both neutrals mentioned in \eqref{flarge} are not very heavy, $M_{a_1},M_{a_2} <350$~GeV, one can hope to  observe two narrow peaks in $b\bar{b}$ production, well separated from each other  in general.

$\triangledown$ If Yukawa sector is similar to Model II (variant (bII)), the eq.~\eqref{YUkII} results in $\chi^b_a\approx 1/\chi^t_a$. In this case the $gg$ partial width can be even larger than the $\bar{b}b$ one.
The cross section $\sigma(gg\to h_a\to \bar{b}b)\approx
\sigma^{(wb)}_{SM}(gg\to h_a\to \bar{b}b)$, and it is difficult to hope for observation of signal of this process  in comparison with background signal $gg\to \bar{b}b$.

{\bf At $\pmb{M_a>350}$~GeV}  the contribution of $h_a\to t\bar{t}$ decay is enlarged so that one can hope to see $h_a$ in $t\bar{t}$ mode (see interesting analysis in \cite{Pilaf14} for particular case of 2HDM-II).

If both neutrals, mentioned in \eqref{flarge} are  heavy, $M_{a_1}>350$~GeV, $M_{a_2}>350$~GeV,
one can hope to  observe either two separated enhancements in $t\bar{t}$ production or even one enhancement (at $|M_{a_1}-M_{a_2}|\le \Gamma_{a_1}+\Gamma_{a_2}$).

\bu {\bf Two photon width}. The widths $h_a\to \gamma\gamma$, $h_a\to Z\gamma$ are described by loop integrals with $W$-loop (contribution is  $\propto\chi^V_a$), $t$-loop (contribution is  $\propto\chi^t_a$), and $H^+$ loop (contribution is  $\propto\chi^\pm_a $). The knowledge of all masses and couplings $\chi^V_a$ don't limit values of $\Gamma(h_a\to \ggam)$ even in 2HDM \cite{GKan}.

As it was found for the simple SM-like scenario ($\chi^t_a\approx 1$) for 2HDM in \cite{GKO}, at $\chi^\pm_1\approx 1$ the contribution of the charged Higgs loop into $\Gamma(h_1\to \gamma\gamma)$ reduces the mentioned width by about 10\% , that is within modern inaccuracy of data. One can also realized SM-like scenario with $\chi^t_a\approx -1$. In this case two photon width can be enhanced by factor $2\div 2.5$ vs SM value \cite{GKO}, which contradicts to modern data. However variation of $\chi^\pm_1$ can reduce this enhancement (see modern studies in  \cite{wrongsign}).

\bu Many results obtained in recent studies can be treated as examples of discussed general picture for separate sets of parameters and particular models.    Our discussion shows that  almost negative results of many such studies have the common origin.

Some authors, estimating opportunities of future experiments, in fact either don't take into account realization SM-like scenario or assume only its weak version (see e.g. \cite{ha14-100}). Some of their results  can be treated as too optimistic if it appears that the  strong SM-like scenario  is realized indeed.

%%%%%%%%%%%%%%%%%%%%%%%%%%%%%
\subsection{Using of charged Higgses, etc.}\label{seccharge}
%%%%%%%%%%%%%%%%%%%%%%%%%%%%%%%%

Here we limit ourself by the group of models with single charged Higgs boson $H^\pm$ (models $2HDM +p_2(HS_cM)+p_1(HS_rM)$).
The discovery of this charged Higgs boson  and the study of its decays should be discussed separately. Below we assume that the mass $M^\pm$ is not extremely large and these particles have good signature.

In the SM-like scenario \eqref{SMlike} the Sum Rules (\ref{SRch})
shows that the coupling $H^\pm W^\mp h_1$ is weak while the couplings $H^\pm W^\mp h_a$ with $a\ge 2$ are close to possible maximal value \eqref{chWa}. It gives following results:

\bu The partial width
$\Gamma(H^+\to W^+h_1)$ is small, while at $M^\pm>M_W+M_a$ the partial width
$\Gamma(H^+\to W^+h_a)$ is relatively large ($a\ge 2$) -- see for example  \cite{lightH}.

\bu The production of Higgs boson $h_1$ in association with $H^\pm W^\mp$ or in decay $H^+\to W^+h_1$ is hardly observable.

\bu  The
search for Higgs bosons $h_a$ can be successful in the following channels:

$q_1\bar{q}_2\to H^+h_a$, $q\bar{q}\to W^\mp H^\pm h_a$ at LHC,

$e\gamma\to \nu H^-h_a$, $e^+e^-\to H^\pm W^\mp h_a$ at $e^+e^-$ collider,

 $\gamma\gamma\to H^\pm W^\mp h_a$ at PLC (Photon Collider).

Some calculations of this type for special variant of 2HDM can be found in \cite{Persy}.
Certainly, $e^+e^-$ collider and PLC have advantages due to much better background conditions.

%%%%%%%%%%%%%%%%%%%%%%
\section{Some effects in the general 2HDM}\label{sec2HDM}
%%%%%%%%%%%%%%%%%%%%%

The most general 2HDM is described by potential with 14 parameters. It contains 2 fundamental fields $\phi_1$, $\phi_2$. The  unitary transformation from these  fields to the fields\lb
$\phi'_1=a_{11}\phi_1+a_{12}\phi_2$, $\phi'_2=a_{21}\phi_1+a_{22}\phi_2$  with corresponding transformation of parameters describe the same physical reality. This transformation is described by 3 parameters. Therefore, total number of physical parameters of model is 11 (see, e.g. \cite{GK05}). Many phenomenological analyses of model deal with its variants, fixed by  some relations among all 14 parameters. Instead of these, it is suggested \cite{GKan} to  describe physical phenomena via the set of 11 well measurable  parameters: $v=246$~GeV -- v.e.v. of Higgs field, masses of 3 neutral Higgs bosons $h_{1,2,3}$ with masses $M_{1,2,3}$ and mass of charged Higgs bosons $M^\pm$, two of three couplings $\chi^V_a$, couplings $g(H^+H^-h_a)$ and $g(H^+H^-H^+H^-)$. There are no theoretical limitations for possible values of these parameters, except limitation for couplings $\chi_a^V$ given by the Sum Rules \eqref{SRV} and general limitations like positivity of potential, etc.

If subsequent observation at LHC  supports SM-like scenario for $V$ and $t$ with reasonable accuracy, even  the data of nearest future on Higgs two photon width will give also limitation for coupling $g(H^+H^-h_1)$ \cite{Ginhhh}.

%%%%%%%%%%%%%%%%%%%%
\subsection{Couplings $Zh_ah_b$}\label{secZhh}
%%%%%%%%%%%%%%%%%%%%%

As it was shown in \cite{GKan}, we have in 2HDM
\be
\chi_{(ab)}^{Z} = \dfrac{g(Z h_a h_b)}{M_Z/v}=-\,\vep_{abc}\chi^V_c\,.
\label{SRZab}
\ee

In the SM-like scenario \eqref{SMlike} it means that among couplings $Zh_ah_b$ only the coupling $Zh_2h_3$ is not small. Therefore the
search for Higgs bosons $h_a, \, h_b$ can be successful in the processes $q\bar{q}\to h_2h_3$ at LHC; $e^+e^-\to h_2h_3$ at \epe\ collider. This opportunity is explored, in particular, for the special variant 2HDM -- Inert doublet model in \cite{Inertbaryo}.

The cross sections of similar processes with production $h_1h_2$ or $h_1h_3$ are negligibly small.

%%%%%%%%%%%%%%%%%%%%%%
\subsection{Triple Higgs production \cite{Ginhhh}}\label{sechhh}
%%%%%%%%%%%%%%%%%%%%%

The measuring of $g(h_1h_1h_1)$ is scheduled in the LHC and other colliders, having in mind two goals.
\\
$\lozenge$ These observations should verify SM into one more point.\\
$\lozenge$ These observations could give information about other Higgs bosons even if we can not observe them in the nearest future.

The accuracy of these measurements can not be high,  since in each case corresponding experiments deal with interference of two channels with identical final state -- an independent production of two Higgses and production of Higgses via $h_1h_1h_1$ vertex (at LHC  -- from $t$-loop). This interference is mainly destructive \cite{hhhinterf}. For example, for 100~TeV hadron collider with total luminosity 3/{\it ab} one can hope to reach accuracy  40\% in the extraction of this vertex \cite{hhhestim}.

According to \cite{GKan}, in the most general 2HDM the mentioned triple vertex is expressed via measurable quantities (factor $M_1^2/v$ is the SM result, quantity $\chi^{hhh}_{111}$ is the relative coupling)\fn{For some particular variant of MSSM the value of triple Higgs coupling with radiative correction $g^{ren}(h_1h_1h_1)$ looks essentially different from its tree form of SM, $M_1^2/v$ \cite{hhhRC}. However, in this very approximation one should take into account renormalization of mass $M_1\to M_1^{ren}$. One can check that this corrected value $g^{ren}(h_1h_1h_1)\approx (M_1^{ren})^2/v$ -- as in  SM \cite{Boudhhh}.}
\bear{c}
g(h_1h_1h_1)=\fr{M_1^2}{v}\chi_{111};\quad%\\
\chi_{111}=
\chi^V_1\left\{1+\!\left(1-(\chi^V_1)^2\right)\left[1
+\sum\limits_c 2\fr{M_c^2}{M_1^2}(\chi^V_c)^2\right]\right.+\\[2mm]
\left.+\!\left(1-(\chi^V_1)^2\right)\fr{2M_\pm^2}{M_1^2}\left[\sum\limits_c\chi^V_c\chi^\pm_c-1+Re\left(\sum\limits_c
\chi^{H^+W^-}_c\chi^\pm_c\fr{\chi^{H^+W^-}_1}{\chi^V_1}\right)\right]\right\}\,.
\eear{triple1-c}

In the SM-like scenario \eqref{SMlike} with $\vep_V=|1-\chi_V^1|\ll 1$ it is easy to estimate
\bear{c}
\chi_{111}\approx
(1-\vep_V/2)\left\{1+\vep_V\left[3+B\vep_V +2B_\pm\left(\chi_1^\pm -1 +\vep_V K_\pm\right)\right]\right\}\,,\\[2mm]
B\sim \sum\limits_c M_c^2/M_1^2; \quad B_\pm=2M_\pm^2/M_1^2\,,\quad K_\pm \sim \chi_c^\pm\,,(b=2,\,3)\,.
\eear{hhhappr}
We see that at  moderate values of parameters, relative coupling $\chi_{111}$ is close to 1, and it is difficult to expect sizable effect\fn{For the particular CP conserving case and with moderate values of parameters such conclusion was obtained in \cite{hhhviol}, \cite{hhhviol1} (see also \cite{hhhMSSMn} for  the CP conserving MSSM).  For the nMSSM (2HDM +Higgs singlet) values $\chi_{111}$ can vary from  1.9 to -1.1 \cite{hhhMSSMn}.}.

\bu \ Special opportunity appears in the SM scenario at $M_2>250$~GeV if $|\chi^t_2|>1$. In this case Higgs boson $h_2$ is relatively narrow and the cross section of gluon fusion $gg\to h_2$ can be  larger than that for the would-be SM Higgs boson with mass $M_2$. In this case process $gg\to h_2\to h_1h_1$ can be seen as resonant production of $h_1h_1$ pair. It allow in principle to discover $h_2$ at LHC. Similar opportunity is absent at $e^+e^-$ colliders.

%%%%%%%%%%%%%%%%%%%%%%%%%
\section{Summary}\label{secsum}
%%%%%%%%%%%%%%%%%%%%%%%%%%%

The big class of Higgs models with arbitrary Yukawa sector obeys simple Sum Rules \eqref{SRV}-\eqref{SRch}, helpful for analysis of future experiments with searching for phenomena beyond SM.

Assuming that the SM-like scenario is realized, for wide class of Higgs models
many now discussed ways to observe additional neutral scalars (new Higgses) are difficult (or inaccessible).

The production of $h_a$ together with charged Higgses looks as the most perspective approach.

One can consider the opportunity to find scalars $h_a$ with mass $M_a<350$~GeV as a relatively narrow peaks in $b\bar{b}$ and (or) $h_1h_1$ channels.

Results of $h_1h_1$ production may differ significantly from the predictions of the SM only at very large values of $M_\pm$ and (or) of vertex $g(H^+H^-h_1)$.

We hope that similar picture is realized in many other models.

\section*{Acknowledgment}

 I am thankful my co-authors in some of presented studies K. Kanishev and M. Krawczyk. The discussions with F. Boudjema, I.Ivanov, A. Maslennikov,  P.Osland, Yu. Tikhonov, M. Vysotsky were useful. I thank S.Munir for comments to the first version of paper.
This work was supported in part by grants RFBR  15-02-05868, NSh-3802.2012.2 and  NCN OPUS 2012/05/B/ST2/03306 (2012-2016).

%\vspace{-0.5em}

\end{document}